\documentstyle[11pt,nyas]{article}

\onecolumn

\input{epsf}

\def\and  {{\it {et al.} }}

\newcommand{\hS}{\hat{S}}
\newcommand{\tw}{\tilde{w}}

\newcommand{\avg}[1]{\langle{#1}\rangle}

\newcommand{\ltsima}{$\; \buildrel < \over \sim \;$}
\newcommand{\lsim}{\lower.5ex\hbox{\ltsima}}
\newcommand{\gtsima}{$\; \buildrel > \over \sim \;$}
\newcommand{\gsim}{\lower.5ex\hbox{\gtsima}}

\begin{document}

\title{Cosmic Statistics of Statistics: $N$-point Correlations}
\ \\
Istv\'an Szapudi\\
\ \\
Canadian Institute for Theoretical Astrophysics\\
University of Toronto\\
60 St. George Street\\
Toronto, Ontario, M5S 3H8\\
\bigskip

 
\begin{abstract}

The fully general calculation of the cosmic error
on $N$-point correlation functions and related
quantities is presented. More precisely, the
variance caused by the finite volume, discreteness,
and edge effects is determined for {\em any} estimator
which is based on a general function of $N$-tuples, such
as multi-point correlation functions and multi-spectra.
The results are printed explicitly
for the two-point correlation function (or power-spectrum),
and for the three-point correlation (or bispectrum). These
are the most popular statistics in  the study of large scale structure, 
yet, the a general
calculation of their variance has not been performed until now.

\end{abstract}

\section{Introduction}

Astrophysics provides prime examples of random fields, such
as the distribution of galaxies and the fluctuations of 
the Cosmic Microwave Background (CMB). Such random fields can
be characterized statistically by a series of well chosen
estimators. The most popular ones include counts-in-cells, 
$N$-point correlation functions, as well as
statistics derived from them. Indeed, there are mathematical
theorems, which state that, under certain conditions, both
series describe a random process fully.

Our goal in astrophysics is not simply estimating these 
statistics, but to constrain underlying theories. 
This aim necessitates a firm control over
the expected statistical errors from a survey.
The theory of errors for finite surveys,
the ``cosmic statistics of statistics'', is therefore of utmost
importance for practical applications. Such a theory
was formulated in the past for counts in cells statistics.
\cite{sc96}$^,$\cite{css}$^,$\cite{bss}

For the $N$-point correlation
functions, however, to date only partial results are published, such
as calculation of the discreteness effects for the two-point
correlation function\cite{ls}, and for $N$-point
correlation functions for Poisson and multinomial point processes\cite{ssa},  
full calculation for the
two-point function under the hierarchical assumption with
edge effects neglected\cite{bern}, and some results
in Fourier space with various degree of approximations.

The aim of this proceedings is to present the general variance
calculation for $N$-point correlation functions with
all major contributions included, such as discreteness effects,
arising from sampling the underlying random field
with a finite number of galaxies, 
edge effects, due to the geometry of the survey and the corresponding
uneven weighting of $N$-tuples, and finite volume effects, 
caused by fluctuations at and above the scale of a survey.
The underlying technique of calculation, as well as the fully
general results are presented here; specializations
such as power spectrum and bispectrum, 
and approximations, such as weakly non-linear perturbation theory, 
hierarchical assumptions, will be presented elsewhere in
collaboration with S. Colombi, and A.S. Szalay\cite{scs}.

The next section sets up the general mathematical framework for
the calculation using computer algebra packages. \S 3 presents
the results for $N = 2$, and $N = 3$. The final discussion
section outlines how the formulae can be used in practice,
as well as describes developments in the immediate future.

\section{General Framework}

Many interesting statistics, such as the
$N$-point correlation functions and their Fourier analogs, can be
formulated as functions over $N$ points in a catalog. The
covariance of a pair of such estimators will be calculated for
a general point process under the assumption that the average
density is {\em a priori} known. This is the obvious generalization
of the  Poisson process when arbitrarily high order correlations
are present. The number of objects is thus varied 
corresponding to a grand canonical ensemble in statistical
physics. The following calculations lean heavily on the elegant
formalism by Ripley\cite{ripley88}, which can be consulted for
details, and are the direct generalization of the 
framework set up by Szapudi \& Szalay.\cite{ssa}

Let $D$ be a catalog of data points to be analyzed, and $R$ randomly
generated over the same area, with averages $\lambda$, and $\rho$
respectively.  The role of $R$ is to perform a Monte Carlo integration
compensating for edge effects, therefore eventually the limit $\rho
\rightarrow \infty$ will be taken. $\lambda$ on the other hand is
assumed to be externally estimated with arbitrary precision. 
No other assumption is taken about the point process. In practice,
$\lambda$ is usually estimated from the same survey, which gives
rise to additional correlations, the ``integral constraint bias''.
This effect will be investigated in more detail elsewhere.

Following\cite{ssa}, let us define symbolically an 
estimator $D^pR^q$, with $p+q = N$ for a 
function $\Phi$ symmetric in its arguments
  \begin{equation}
	D^pR^q = \sum\Phi(x_1,\ldots,x_p,y_1,\ldots,y_q),
  \end{equation}
with $x_i \ne x_j \in D, y_i \ne y_j \in R$. As an example, the
two point correlation function corresponds to
 $\Phi(x,y) = [x,y \in D, r \le d(x,y) \le r+dr]$, 
where $d(x,y)$ is the distance between the two points, and
$[condition]$ equals $1$ when $condition$ holds, $0$ otherwise.
Ensemble averages can be estimated via factorial moment
measures, $\nu_s$.\cite{dv72}$^,$\cite{ripley88} 
In the Poisson limit $\nu_s = \lambda^s\mu_{s}$, 
where $\mu_s$ is the $s$ dimensional Lebesgue measure,
and in the most general case $\nu_s f(x_1,\ldots,x_s)\lambda^s\mu_s$.
The function $\lambda^sf(x_1,\ldots,x_s) = F(x_1,\ldots,x_s)$ 
can be identified as the full, i.e. non-reduced, $N$-point correlation
function. The connection between these and the reduced $N$-point
correlation functions is well known\cite{ssa}, and can
be obtained through either generating functions, or recursions.

The general covariance of a pair of estimators is
  \begin{equation}
	E(p_1,p_2,N_1,N_2) = 
        \avg{\hat D_a^{p_1}\hat R_a^{q_1}\hat D_b^{p_2}\hat R_b^{q_2}} = 
	\sum_{i,j}{p_1 \choose i}{p_2 \choose i} i!
	{q_1 \choose j}{q_2 \choose j} j!\,
	S_{i+j}\lambda^{-i}\rho^{-j},
  \label{eq:cov}
  \end{equation}
with $p_1+q_1 = N_1, p_2 + q_2 = N_2$, $S$ will be
specified later, $\hat{\,} $ denotes
normalization with $\lambda, \rho$ respectively, i.e.
$(\hat D = D/\lambda, \hat R = R/\rho$.
The expression simply describes the fact that out of the $p_1$ and $p_2$
different data points in $D$ we have an $i$-fold degeneracy, as well
as a $j$-fold degeneracy in the random points drawn from $R$. 
To simplify the exposition of the 
calculation, it is convenient to assume from the 
very beginning the $\rho\rightarrow\infty$, i.e.
the random process employed for the Monte Carlo
integration of the shape of the survey is arbitrarily
dense. This is usually achievable in practice, thus it
will not be considered further. The above equation describes the
cross-correlation of two estimators even for two
different objects as well: e.g., two particular bins of the
two- and three-point correlation functions. 
An interesting special case, $N_1 = 1$ (the average density
in the survey)  and $N_2 \ge 2$, is needed for calculating the
``integral constraint'' correction. 

When the random process is arbitrarily dense only $j = 0$ survives,
  \begin{eqnarray}
	&& E(p_1,p_2,N_1,N_2) = 
        (\avg{\hat D_a^{p_1}\hat R_a^{q_1}\hat D_b^{p_2}\hat R_b^{q_2}} = 
        \\ && 
	\sum_{i,j}{p_1 \choose i}{p_2 \choose i} i!\lambda^{-i}
	\hS_{i} f(1,2,\ldots,p_1,N_1+1,\ldots, N_1+p_2-i), \nonumber
  \label{eq:cov1}
  \end{eqnarray}
where $\hS$ is now an operator acting on $f$,
  \begin{equation}
	\hS_{k} = \int\Phi_a(1\ldots N_1)
              \Phi_b(1\ldots i, N_1+1, \ldots, N_1+N_2-i)
              \ldots
              \mu_{2N-k}.
  \end{equation} 
The operator $\hS_k$ is analogous to the phase space integral
$S_k$\cite{ssa}. The dot emphasizes that the integral can be
performed only after $\hS_k$ is acted on $f$ which is part of the
measure. The phase space has to be calculated in the general
case via the full factorial moment measure of which $f$ is an
integral part.
Throughout the paper we use the convention that ${k\choose l}$ is nonzero
only for $k \ge 0, l\ge 0$, and $ k \ge l$, and the variables $x_i$
are denoted with $i$ for simplicity.
Here $\Phi_a$ and $\Phi_b$ denote two different functions, for instance
corresponding to two radial bins of two estimators of the same or
different orders. In the above formula the symmetry of $\Phi$ in
its arguments was heavily relied on to achieve the above
``standard'' representation of the operator.

The dependence of $S_k$ on $a,b$, and $N$ is not noted for 
convenience\cite{ssa}, but they will be
assumed throughout the paper.  The estimator\cite{ssa} for the generalized
$N$-point correlation function is $(\hat D-\hat R)^N$, or
more precisely,
  \begin{equation}
  	\tw_N = \frac{1}{S}  \sum_i 
	{N \choose i}(-)^{N-i} (\frac{D}{\lambda})^i (\frac{R}{\rho})^{N-i},
  \end{equation}
where $S = \int\Phi\mu_{N}$ (without subscript).  In this case
$S$ is a number since it corresponds to the Poisson catalog with
its simple factorial moment measure.
The average of the estimator yields
  \begin{eqnarray}
  	&& \avg{\tw_N} = \frac{1}{S}  \sum_i 
	{N \choose i}(-)^{N-i} \int \Phi(1,\ldots,N)f(1,\ldots,N)\mu_N = \\
        && \frac{1}{S}\int\Phi(1,\ldots,N)\xi_N(1,\ldots,N)\mu_N. \nonumber
  \end{eqnarray}
Since the role of $\Phi$ is effectively a window, with a 
window operator $\hat{W}$ this can be written symbolically
as $\avg{\tw_N} = \hat{W}\xi_N/\hat{W}$.
The asymptotic centered 
covariance between two estimators of the above for a general
point process in the limit of $\rho\rightarrow\infty$ can
be written according to the previous considerations as
\begin{eqnarray}
 &&  \avg{\delta\tw_{N_1}\delta\tw_{N_2}} =  
  \avg{\tw_{a,N_1}\tw_{b,N_2}} - \avg{\tw_{a,N_1}}\avg{\tw_{b,N_2}} = \\
 && \frac{1}{S^2}\sum_{i,j}{N_1 \choose i}{N_2 \choose j}(-)^{i+j}
  \left[E(i,j,N_1,N_2)-\hS_0f(1,\ldots,i)f(N_1+1,\ldots, N_1+j)\right].
 \nonumber
\end{eqnarray}
In the above the operator $S_0$ acts on the multiple of the
two $f$'s on the right. The above equation is the main
result of this paper. While it is quite cumbersome, it is 
easily expandable with the help of computer algebra, as demonstrated
by the examples of the next section. The special cases rendered
will also illustrate how the simplicity of the proposed
class of estimators exactly manifests itself by a mass cancellation
of terms. Any other estimator would have extra terms in the
variance.\cite{ssa}

\section{The cosmic error on the two-and three-point correlation
functions}

The above equation was entered into a computer algebra package.
The symmetries and simplicity of the estimator 
give rise to cancellations and possibilities for collecting
similar terms. This is the reason why the final result
for the two-point correlation function has only three to six terms, 
while formally it could have up to about a hundred.
Alternative estimators,
such as $DD/RR-1$, etc. would not yield significantly
less cancellations, therefore error-calculation for them
was not attempted; although the same formalism applies.

\subsection{The two-point function} 

The covariance for the two-point function (or any quantity
estimated from doublets of points, such as the power spectrum)
can be expressed after the cancellations and the possible
simplifications as
\begin{eqnarray}
  \label{eq:twopoint}
  \avg{\delta\tw_2^a\delta\tw_2^b} &&= \frac{1}{S^2}\Bigr\lbrace\Bigr. 
  \int\Phi_a(1,2)\Phi_b(3,4)
   \left[ \xi_4(1,2,3,4)+2\xi(1,3)\xi(2,4)\right] + \\
  && \frac{4}{\lambda}\int\Phi_a(1,2)\Phi_b(1,3)
   \left[ \xi(2,3)+\xi_3(1,2,3)\right] + \nonumber \\
  && \frac{2}{\lambda^2}\int\Phi_a(1,2)\Phi_b(1,2)
   \left[ 1+\xi(1,2)\right] \Bigl.\Bigr\rbrace \nonumber .
\end{eqnarray}       
The above equation is essentially identical to the result
of Hamilton\cite{h93} where he calculates the variance of $\delta$, the
fluctuation field itself. This is not at all surprising.\cite{ssa}
The estimator contains exactly the same terms and
coefficients (regardless of the choice of $\Phi$) as $\delta$
itself, which strongly suggests that it is (nearly) optimal.

The above formula contains the different contributions to the
error\cite{sc96} entirely mixed. Approximate separation of the
different terms appears to be fruitless.
The only general point to be
made is that discreteness effects are absent in the first
term, while they are present (mixed with the other two
effects) in the $1/\lambda^s$ terms, with $s>0$. This is
naturally true for the higher order calculations as well.

It is worth to emphasize again, that this 
formula applies for the generalized $2$-point function, including
the ``traditional'' $2$-point function, and any of its incarnations,
such as the  the power spectrum. In the latter case, 
esthetic or practical reasons might dictate 
that the errors on the power spectrum are expressed
in terms of the power-spectrum, bi-, and tri-spectrum, instead
of the two-, three-, and four-point correlation functions.
Since the connection
is a simple Fourier transform, this trivial exercise is left
for the reader. Explicit formulae, aimed at practical applications
for power-spectrum will be presented elsewhere.\cite{scs}

\subsection{The three-point correlation function}
The same method yields (co)variance for higher order estimators
as well. We present another example, the generalized
three-point correlation function. Its variance, using
again the main result of the proceeding, 
translates into:
\begin{eqnarray}
  \avg{\delta\tw_3^a\delta\tw_3^b} &&= \frac{1}{S^2}\Bigl\{\Bigr. \\
  && \int\Phi_a(1,2,3)\Phi_b(4,5,6)\bigl[\bigr. \xi(1,2,3,4,5,6)+ \nonumber \\
  &&3\xi(1,2)\xi(3,4,5,6) +9\xi(1,4)\xi(2,3,5,6)+ \nonumber \\
  &&3\xi(4,5)\xi(1,2,3,6) +9\xi(1,5,6)\xi(2,3,4)+ \nonumber \\
  &&9\xi(1,4)\xi(2,3)\xi(5,6) +6\xi(1,4)\xi(2,5)\xi(3,6)\bigl.\bigr] 
  \nonumber \\
  &&\frac{9}{\lambda}\int\Phi_a(1,2,3)\Phi_b(1,4,5)
     \bigl[\bigr. \xi(1,2,3,4,5)+ \nonumber \\
  &&\xi(2,3,4,5) +2\xi(1,2)\xi(3,4,5)+ \nonumber \\
  &&2\xi(1,4)\xi(2,3,5) +\xi(2,3)\xi(1,4,5)+ \nonumber \\
  &&4\xi(2,5)\xi(1,3,4) +\xi(4,5)\xi(1,2,3)+ \nonumber \\
  &&\xi(2,3)\xi(4,5)+2\xi(2,4)\xi(3,5)\bigl.\bigr] +\nonumber \\
  &&\frac{18}{\lambda^2}\int\Phi_a(1,2,3)\Phi_b(1,2,4)
     \bigl[\bigr. \xi(1,2,3,4)+ \nonumber \\
  &&2\xi(1,3,4) +\xi(1,2)\xi(3,4)+ \nonumber \\
  &&2\xi(1,3)\xi(2,4) + \xi(3,4) \bigl.\bigr] \nonumber \\
  &&\frac{6}{\lambda^3}\int\Phi_a(1,2,3)\Phi_b(1,2,3)
     \bigl[\bigr. \xi(1,2,3)+ 
  3\xi(1,2) +1\bigl.\bigr] \Bigl.\Bigr\}. \nonumber
\end{eqnarray}       
For simplicity, in the above formula the order of $\xi$ is notated
with the number of arguments only, e.g., $\xi_3(1,2,3) = \xi(1,2,3)$.
The above equation is is less obviously useful then that of the
two-point correlation function. Nevertheless, given a model
for the higher order correlation functions, such as weakly
non-linear perturbation theory, or any well specified version
of the hierarchical assumption, the equation can easily be turned
into a practical computer program.\cite{scs}

The variance of the four-point and higher order correlation functions 
can be calculated  analogously, but it would make no sense to print 
the results.  When needed, the formulae
generated by computer algebra can be transformed into Fortran or
C-code directly.

\section{Discussion}

The above method, and the explicit formulae given, can be used
to evaluate the cosmic error on any statistical measure based
on $N$-tuples of a distribution. This includes, but is not
limited to, $N$-point correlation functions, $N$-th order 
cumulants, cumulant correlators, multi-spectra, etc.

The above calculation was performed only for the
best $N$-point estimator.\cite{ssa}
Any other estimator can be calculated analogously, but
be warned that the resulting number of terms can be
overwhelming due to the insufficient cancellation arising from
suboptimal edge correction.

The fact that the average density is assumed to be given as
an outside parameter appears to be fairly restrictive.
However, maximum likelihood context, which is
the most important potential practical application of the
results, it is easy to remedy the situation. The proposed estimator\cite{ssa}
can be trivially changed by {\em not} normalizing with
the average density $\lambda$. This introduces only a small
modification to the final formulae, and a set of estimators,
including that of the the average count, contains all information
need for constructing likelihood function.
Such a procedure yields full statistical description,
takes into account fluctuations in the mean, and the
fact the average is estimated from the same surveys (``integral
constraint''). Practical demonstration of this procedure
will be presented elsewhere.\cite{scs}

The proposed estimator used here is not connected for $N \ge 4$.\cite{ssa}
Therefore the  calculations for the  higher order connected
estimator should be modified for accurate results for the
connected moments. This trivial but tedious generalization
is left for future research.

The explicit formulae can be specialized for several cases, 
which will be presented elsewhere.\cite{scs} The interesting limits
include Poisson, Gaussian, weakly-nonlinear, strong
correlations, hierarchy, shot noise limited, continuous limit 
etc. The detailed discussion of these limits, and specializations
to particular statistics, such as $N-$-point correlation
functions, multi-spectra, would go beyond the scope of the
present exposition. Similarly, the main equation yields
cross correlations between different statistics as well,
a must for any investigation in the maximum likelihood framework.

Finally, it is worth to note here, that recent advances
in algorithms for calculating $N$-point correlation functions
render these objects more interesting then ever. Fast
algorithms\cite{cnms} will make it possible
to calculate $N$-point functions from very large catalogs,
be it artificial or real data, which undoubtedly will culminate
in new insights into the subject.
The formulae presented in this proceedings will provide the firm 
theoretical grounding for any such investigation.

{\bf Acknowledgments}

This work will form the integral part of a project in 
collaboration with S. Colombi, and A.S. Szalay soon to be
published.\cite{scs} 

\def\apj { ApJ\, }
\def\aap {A \& A\, }
\def\ajs{ ApJS\, }
\def\apjs{ ApJS\, }
\def\mnras { MNRAS\, }
\def\apjl { Ap. J. Let.}



\section{An Alternative Technique}

An alternative method of calculation is possible, which is instructive
and insightful, even if less rigorous mathematically then the above formalism
using factorial moment measures. This alternative technique is well
suited for obtaining quick results for low order estimators by hand.
We demonstrate the calculation
for two-point correlation function, higher order results can be obtained
analogously, although it quickly becomes prohibitevely tedious.

Let us assume that the survey is divided into $K$ bins, each of
them with fluctuations $\delta_i$, with $i$ running from $1\ldots K$.
For this configuration our estimator can be written as
\begin{equation}
   \tilde w = f_{12}\delta_1\delta_2.
\end{equation} 
The above equation uses a ``shorthand'' Einstein convention:
$1,2$ substituting for $i_1,i_2$, and repeated indices summed.
The pairwise weights $f_{12}$ correspond to $\Phi$ in the
main body of the paper, and it is assumed that the two indices cannot
overlap.

The ensemble average of the above estimator is clearly
$f_{12}\xi_{12}$. The continuum limit (co)variance between
bins $a$ and $b$ can be calculated
by taking the square of the above, and taking the ensemble average.
 \begin{equation}
   \avg{\delta \tilde w^a \delta \tilde w^b} = f_{12}^af^b_{34}\left(
   \avg{\delta_1\delta_2\delta_3\delta_4}-
   \avg{\delta_1\delta_2}\avg{\delta_3\delta_4}\right).
  \end{equation} 
Note that the averages in this formula are not connected moments, 
which are distinguished by $\avg{}_c$. 

The above equation yields only the continuum limit terms. To add
Poisson noise contribution to the error, note 
that it arises from the possible overlaps between the indices
(indices between two pairweights {\em can} still overlap!).
In the spirit of infinitesimal Poisson models, we replace
each overlap with a $1/\lambda$ term, and express the
results in terms of connected moments. There are three
possibilities, i) no overlap (continuum limit)
\begin{equation}
   f_{12}^af^b_{34}
    \left(\xi_{1234}+\xi_{13}\xi_{24}+\xi_{14}\xi_{23}\right),
\end{equation} 
ii) one overlap (4 possibilities)
 \begin{equation}
    \frac{4}{\lambda}f_{12}^af^b_{13} \left(\xi_{123}+\xi_{23}\right),
  \end{equation} 
iii) two overlaps (2 possibilities)
 \begin{equation}
    \frac{2}{\lambda^2}f_{12}^af^b_{12} \left(1+\xi_{12}\right),
  \end{equation} 
In these equations, for the sake of the Einstein convention
we used $\xi(i,j,k,l) \rightarrow \xi_{ijkl}$. The above
substitutions (rigorously true only in the infinitesimal Poisson
sampling limit) become increasingly accurate with decreasing
cell size. In that limit, adding the above three equations is 
equivalent to Eq.~\ref{eq:twopoint}. 

\end{document}